\documentclass[12pt,journal,draftcls,letterpaper,onecolumn]{IEEEtran}

\usepackage{graphicx} 
\usepackage{amsmath}
\usepackage{amssymb}
\usepackage{url}
\newcommand{\comment}[1]{}

\usepackage{amsthm}

\newcommand{\vx}{\mbox{${\bf x}$}}

\newcommand{\ve}{\mbox{${\bf e}$}}

\newcommand{\vzero}{\mbox{${\bf 0}$}}
\newcommand{\vone}{\mbox{${\bf 1}$}}

\newcommand{\mA}{\hbox{{\bf A}}}

\newcommand{\mI}{\hbox{{\bf I}}}

\newcommand{\mL}{\hbox{{\bf L}}}
\newcommand{\mP}{\hbox{{\bf P}}}

\newcommand{\mW}{\hbox{{\bf W}}}

\newcommand{\ga}{\alpha}

\newcommand{\gd}{\delta}

\newcommand{\gm}{\mu}

\newcommand{\gr}{\rho}

\newcommand{\gt}{\tau}

\newtheorem{theorem}{Theorem}[section]

\newtheorem{lemma}{Lemma}[section]

\theoremstyle{remark}
\newtheorem{claim}{Claim}
\theoremstyle{remark}
\newtheorem{example}{Example}
\newcommand{\E}{\mathbb{E}}

\newtheorem{coro}[section]{Corollary}

\newcommand{\beq}{\begin{equation}}
\newcommand{\eeq}{\end{equation}}
\newcommand{\bea}{\begin{array}}
\newcommand{\ena}{\end{array}}
\newcommand{\bds}{\begin {itemize}}
\newcommand{\eds}{\end {itemize}}
\newcommand{\blm}{\begin{lemma}}
\newcommand{\elm}{\end{lemma}}
\newcommand{\bthm}{\begin{theorem}}
\newcommand{\ethm}{\end{theorem}}
\newcommand{\bcl}{\begin{claim}}
\newcommand{\ecl}{\end{claim}}
\newcommand{\bcr}{\begin{coro}}
\newcommand{\ecr}{\end{coro}}

\newcommand{\rarrow}{{\rightarrow}}

\begin{document}

 \title{Boundary value problems in consensus networks}

\author{Amir Leshem \thanks{A. Leshem is with the Faculty of Engineering,
        Bar-Ilan University, 52900, Ramat-Gan. Email: leshema@eng.biu.ac.il. The work of A. Leshem is partially supported by ISF grant 903$\backslash$2013, Signal processing for resource limited sensor networks.}, {\em Senior member}, Maziyar Hamdi, and Vikram Krishnamurthy\thanks{Maziyar Hamdi and Vikram Krishnamurthy are with Dept. of Electrical and Computer  Engineering, University of British Columbia, Vancouver, Email: \{maziyarh,vikramk\}@ece.ubc.ca.}, {\em Fellow}, {\em IEEE}.}
\date{}
\renewcommand{\baselinestretch}{2}
 \maketitle

\begin{abstract}
This paper studies the effect of boundary value conditions on consensus networks. Consider a network where some nodes keep their estimates constant while  other nodes average their estimates
with that of their neighbors. We analyze such networks and show that in contrast to standard consensus networks,  the network estimate converges to a general harmonic function on the graph. Furthermore, the final value depends only on the value at the boundary nodes. This has important implications in consensus networks -- for example, we show that consensus networks are extremely sensitive to the existence of a single malicious node or consistent errors in a single node. 
We investigate the existence of boundary nodes in human social networks via an experimental study involving human subjects. Finally, the paper is concluded with the numerical studies of the boundary value problems in consensus networks.
\end{abstract}

\section{Introduction}
\label{sec:intro}


In the last decade, many papers have   studied  consensus in networks, see  \cite{boyd2005gossip,boyd2006randomized,consensus,freeman2006stability,kempe2003gossip,kempe2004spatial} for some early work on consensus formation and gossip algorithms. These and subsequent papers concentrate on various algorithms to achieve consensus in a network. Typical consensus averaging algorithm is composed of a distributed averaging where each node in the network averages its data with its neighbors using a (possibly time varying) weighting. The simplest algorithms to analyze involve fixed averaging where the weighting matrix in the graph is doubly stochastic. A more sophisticated approach involves gossip based algorithms in which  each node randomly picks a neighbor (with the possibility of not choosing any) and updates its data with the neighbor, and updates the weighting as well. It is shown that this scheme also converges to the population average.
A significant amount of research concentrated on the rate of convergence to the consensus using analysis of the second eigenvlue of the network weighting matrix, on the design of such matrices and on analysis of distributed convergence to good weighting which accelerates the convergence \cite{olshevsky2009convergence}. Similar results are also available for the random gossip model, although the analysis is more intricate \cite{boyd2006randomized} due to the stochastic nature of the gossip process.
Much work has also been devoted to the analysis of quantization schemes \cite{kashyap2007quantized}, and to imperfect channels which introduce noise \cite{rajagopal2011network,kar2009distributed}. The close analogy between diffusion processes and consensus averaging has also been a fruitful analogy which led to many interesting results, see e.g., \cite{sardellitti2010fast,chen2012diffusion,diffusion}.  
More recent results can be found in the review paper by Dimakis et al. \cite{dimakis2010gossip}.

Most papers in consensus analysis
focus on  the case of {\em initial value problems}, where a distributed (possibly time varying) operator is applied to the data.
In contrast, this paper considers {\em boundary value problems} in consensus networks.
Such boundary value problems were originally motivated by  continuous diffusion problems, where  boundary conditions play an important role, and in many cases influence  the nature of the solution. We first analyze the effect of fixed boundary conditions on the network, and show that even when a single node is subject to a boundary condition, it is the value of the data in this node that actually determines the limiting value. For simplicity we focus on the constant averaging matrix. However in an extension of this work we analyze the effect of boundary value conditions on randomized gossip algorithms. Not surprisingly, our study of boundary conditions leads to the analysis of general harmonic functions on graphs. A good overview of such functions, which also play an important role in analyzing random walks on graphs is given by Benjamini and Lovasz \cite{benjamini2003harmonic}. The paper by Bendito et al. \cite{bendito2000solving} consider the solution of discrete boundary value problems using Green's functions. The solution is quite complicated and in this paper we take a different path, which leads to closed form solution and better intuitions regarding the solution.

The main results of the paper determine the asymptotic value of the nodes in a consensus network, subject to boundary conditions. We show that in this case the value is no longer constant at all nodes, but consists of a general harmonic function on the graph. We then prove an analog of the theorem that an harmonic function with constant value on the boundary is constant, and conclude that even fixing a single node can be used to drive the network to any desired consensus. This fact has several important consequences: Distributed detection in consensus networks can be significantly harmed even in the presence of a single malicious node. This also has impact on the sensitivity to consistent errors.

Wang and Krim \cite{wang2013control,wang2014analysis} have recently discussed the control of beliefs in social networks. They studied the maximal change in beliefs as a function of the degree distribution of the network. In this paper, we show that the beliefs can be steered towards any constant value by impacting a single node. Furthermore, any given change in the beliefs can be implemented by carefully selecting a small number of nodes to influence and using a time varying periodic strategy. The choice of nodes to impact has mainly an effect on the speed of convergence. to achieve this we analyze network with periodic boundary conditions.

The applications of this paper go beyond the obvious networking setups. Interestingly, the results of this paper also generalize the results of Cohen and Peleg \cite{cohen2005convergence}, \cite{cohen2008convergence}, \cite{cohen2008local} on distributed convergence of groups of robots to the center of gravity of the robots. These results are a special case of our main theorem. According to the model Cohen and Peleg each robot moves in the average direction of the robots it sees. This can be considered as the application of two consensus processes for the $x$ and $y$ coordinates of the robots. Furthermore, our results show that when some robots are fixed and serve as beacons for the others, the rest of the group will converge to the center of gravity of the fixed robots.

The structure of the paper is as follows: Section~\ref{problem formulation}  poses the problem of consensus with boundary value conditions. Section~\ref{s:main_theorem} presents the main result of the paper, namely  Theorem~\ref{main_theorem}.
We provide several consequences for specific network topologies.  Section~\ref{gossip}  describes two randomized gossip based versions of the algorithm and prove their convergence. Section~\ref{consequences} discusses consequences to advertising in social networks as well as to distributed decision making in sensor networks. We present an experimental study in section~\ref{experimental} to investigate the existence of boundary nodes in a community. Interestingly, our experiment verified the existence of such boundary nodes (who are not affected by their neighbors) in the population under study (a group of undergraduate students). We discuss the sensitivity of sensors networks to malicious nodes and show that even a single malicious node can practically steer the network towards any desired decision. We also point out that consistent bias in a single node can make the entire network useless. Section~\ref{periodic}  extends the results to time varying periodic boundary conditions and provide closed form to the limit of the network. Finally, numerical examples are presented in Section~\ref{simulations}.

\section{Problem formulation}
\label{problem formulation}
Consider a network with $K$ boundary nodes $1,\ldots,K$ and $M$ internal nodes ${K+1},\ldots,{K+M}$.
The connectivity graph of the network is denoted by $G=\left(V,E \right)$, where $V =\{1,2, \ldots,{K+M}\}$, and $E \subset  V\times V$. Let $\mA$ denote the adjacency (connectivity) matrix of the graph $G$ with elements $a_{i,j}$ given by:
\beq
a_{i,j}=1 \iff \{i,j\} \in E.
\eeq
Define the set of neighbors of node $n$ as  $N_n=\left\{m \in V: \left\{n,m\right\}\in E \right\}$.
The aim of this paper is  to analyze the dynamics of consensus averaging when the network is subject to boundary value constraints, i.e., when the boundary nodes do not average the estimates of their neighbors and keep their estimates constant.

At time $0$, each node $1 \le n \le M+K$  has an initial value $x_n(0)$. Subsequently, each internal node $n>K$ averages the input from its neighbors.
\beq
x_n(t+1)=p_{n,n} x_n(t)+\sum_{\left\{m \in N_n \right\}} p_{n,m}(t)x_m(t),
\eeq
where $0 \leq p_{i,j} \leq 1$ (for $K<i\leq K+M$ and $1\leq j \leq K+M$) are the averaging weights. The system continues until convergence is achieved. In the existence of boundary nodes, the behavior of the network is very different than the behavior of a standard consensus network.
Before proceeding, let us define the following matrices which are used to analyze the behavior of the system:
Let $\mP_i$ be the weights used in averaging internal nodes with elements \beq \mP_i(k,l) = p_{k+K, l+K} \quad 1 \leq k,l \leq M,  \eeq and $\mP_e$ be the weights assigned to the external boundary nodes with elements \beq \mP_e(k,l) = p_{k+K, l} \quad 1 \leq k \leq M, 1 \leq l \leq K.  \eeq
The system dynamics is described by the following equations:
\beq
\bea{lll}
\hbox{Initial conditions: } &  & \\
& \vx_b(0)=[x_1(0),...,x_K(0)]^T, & \\
& \vx_i(0)=[x_{K+1}(0),...,x_{M+K}(0)]^T & \\
\hbox{System dynamics: } & & \\
\hbox{For all $t>0$: } & & \\
& \vx_i(t+1)=\mL \vx(t), &
\ena
\label{boundary_TI}
\eeq
where
\beq
\vx(t)=\left[ \vx_b^T(t),\vx_i^T(t)\right]^T=\left[x_1(t),...,x_{M+K}(t)\right]^T,
\eeq
and
\beq\label{eq:L}
\mL = \left[
\bea{c|c}
\mI_{K\times K} & {\bf 0}_{K\times M} \\
\hline \\
\mP_e & \mP_i
\ena
\right].
\eeq
Here, $A^T$ is used to denote the transpose of  matrix $A$. Note, that in this case the boundary conditions are time invariant. Later in Section~\ref{boundary_periodic}, an analysis is presented for the general case. Also note that $\mL$ is a stochastic matrix, since it averages each $x_n(t)$ with its neighbors. For simplicity, in the next section, we analyze the case of constant averaging matrix. This will provide good insight into the general case of time varying averaging. The generalization to randomized gossip algorithms is investigated in the subsequent section.

\section{Asymptotic Analysis of Boundary Value Problems}
\label{s:main_theorem}
This section presents the
 main result of the paper, namely  the analysis of  system (\ref{boundary_TI}). It is shown that the stable point of (\ref{boundary_TI}) is independent of the values stored at the internal nodes of the network. This has important consequences for consensus based networks. Then, some special cases of boundary conditions are discussed and we show that, in general, the network does not converge to a consensus, unless the boundary conditions are constant functions.

Our first goal is to study the stable points of (\ref{boundary_TI}). The following theorem provides existence and uniqueness of the stable point.
\bthm
Consider the system given in (\ref{boundary_TI}) with $\mL$ defined in (\ref{eq:L}). Assume that the graph $G$ is connected.
and the stochastic matrix $\mL$ describes a Markov chain with absorbing states $1,..,K$ and transient states $K+1,...,N$ (This amount to the fact that all the internal nodes are averaging inputs from nodes $1,...,K$ at some stage). Then the following holds:
\begin{enumerate}
\item $\mI-\mP_i$ is an invertible matrix.
\item
System (\ref{boundary_TI}) always converges to a limit point which is given by:
\beq \label{main_eq}
\bea{lcl}
\vx_\infty(\vx_b(0))&=&\lim_{k \rarrow \infty} \vx(k) \\
                    &=& \left[\vx_b^T, \left(\left( \mI-\mP_i \right)^{-1} \mP_e\vx_b\right)^T \right]^T.
\ena
\eeq
\end{enumerate}
\label{main_theorem}
\ethm
\begin{proof}
To prove the first part, note that $\mP_i$ is a weakly sub-stochastic matrix, in which the sum of all rows is less or equal 1 and there exists at least one row which sums to less than 1. To show this, consider the matrix $\mL$ as the transition probability matrix of a Markov chain. This Markov chain has absorbing states $1,..,K$, and the stationary distribution is concentrating on the absorbing states. Examining the powers of the matrix $\mL^k$, this implies that $\lim_{k \rarrow \infty} \mP_i^k={\bf 0}$. Therefore, the spectral radius of $\mP_i$ satisfies $\gr(\mP_i)<1$ and, thus, $\mI-\mP_i$ is invertible.

Consider now, computing the limiting value of (\ref{boundary_TI}) to prove the second part. It follows via  induction  that for each $k$
\beq
\mL^k = \left[
\bea{c|c}
\mI & {\bf 0} \\
\hline \\
\left(\mI+\mP_i+...+\mP_i^{k-1}\right)\mP_e & \mP_i^k
\ena
\right].
\eeq
To see this, note that at $k=1$ this trivially holds. By induction:
\beq
\mL^{k+1}=\mL^k \mL
\eeq
Using block by block multiplication, it is seen  that the structure is preserved and the left bottom block satisfies:
\beq
\sum_{m=1}^{k}\mP_i^m \mP_e=\left(\sum_{m=1}^{k-1}\mP_i^m \mP_e\right) \mI + \mP_i^{k} \mP_e.
\eeq
Similarly, the right bottom block is $\mP_i^{k+1}$.
Since the spectral radius of $\gr(\mP_i)<1$, it follows that $\mI-\mP_i$ is invertible.
Let
\beq
\mL_{\infty}=\lim_{k \rarrow \infty}\mL^k =
\lim_{k \rarrow \infty}
\left[
\bea{c|c}
\mI & {\bf 0} \\
\hline \\
\left(\mI+\mP_i+...+\mP_i^k\right)\mP_e & \mP_i^k
\ena
\right]
\eeq
Since the spectral radius of $\mP_i$ is less than 1:
\beq
\mL_{\infty}=
\left[
\bea{c|c}
\mI & {\bf 0} \\
\hline \\
\left(\mI-\mP_i\right)^{-1}\mP_e & 0
\ena
\right].
\eeq
\bcl
\label{cl:Linf_stochastic}
$\mL_{\infty}$ is stochastic.
\ecl
For each $k$, $\mL^k$ is a stochastic matrix. Hence, by a simple continuity argument $\mL_{\infty}$ is also stochastic.

Therefore, for any vector $\vx(0)=\left[\vx_b^T(0),\vx_i^T(0) \right]^T$
we obtain that
\beq\label{eq:asymptotic}
\lim_{k \rarrow \infty} \vx(k) = \left[\vx_b^T, \left(\left( \mI-\mP_i \right)^{-1} \mP_e\vx_b\right)^T \right]^T.
\eeq
An alternative way to observe the equilibrium equation is by writing
\beq \label{temp}
\vx_\infty=\mL\vx_\infty.
\eeq
Let $\vx_i^\infty = \lim_{k\rightarrow \infty} \vx_i(k)$. Then, from (\ref{temp})
\beq
\vx_i^\infty = \mP_e \vx_b+\mP_i \vx_i^\infty,
\eeq
which translates into
\beq
\vx_i^\infty=\left(\mI-\mP_i \right)^{-1} \mP_i \vx_b
\eeq
\end{proof}

This result implies that in the existence of boundary nodes, the network's final state does not depend on the values of the internal nodes. This simple observation has significant implications on consensus based networks which will be analyzed  in the subsequent sections.

Here, an interesting property of the limiting function, which has important implications to sensor and social networks, is presented. The following theorem is a
\bthm
Consider the system given in (\ref{boundary_TI}) with $\mL$ defined in (\ref{eq:L}). Assume that the initial values of boundary nodes  $\vx_b(0)=\gm \vone_K$ where $\vone_K$ is a $K$ dimensional vector of $1$'s. Then,
$\vx_\infty(\vx_b(0))=\gm \vone_N$.
\label{thm:constant}
\ethm
\begin{proof}  Recall from Claim~\ref{cl:Linf_stochastic}  that, $\mL_{\infty}$ is stochastic. Therefore, the vector $\vone_N$ is an eigenvector of $\mL_\infty$ with eigenvalue $1$ and  $\left(\mI-\mP_i\right)^{-1}\mP_e \gm \vone_K = \gm\vone_{N-K}$. \end{proof}
The above theorem implies that if the value on the boundary is constant the network will converge to the same value.
This result has significant implications to both social network and decision making in sensor networks which are discussed, in details, in Section~\ref{consequences}.

The operator $\mL$ averages each $x_n(t)$ with its neighbors in the graph. Hence, in steady state, $\vx_\infty(\vx_b(0))$ is a generalized harmonic function on the graph $G$. Furthermore, by the above analysis, the boundary conditions completely determine the final state. It follows that, if the values of the
boundary nodes are not constant, the network will not converge to a consensus at all, as the following example shows:
\begin{example}
Assume that $G$ is a line graph as described in figure \ref{line_graph} where $V=\left\{0,...,N-1 \right\}$ and $\{i,j\}\in E \iff j=i+1$.
and for $0<n<N-1$ and $0<\ga<1$:
\beq
x_n(t+1)=\ga x_{n}(t)+ \frac{1-\ga}{2} \left(x_{n-1}(t)+x_{n+1}(t)\right).
\label{line_dynamics}
\eeq
Assume that the boundary nodes are $0,N-1$ and that the boundary values are $x_0(t)=a, x_{N-1}(t)=b$ and $a \neq b$. The network converges to a linear function where $x_n^{\infty}=a+\frac{n(b-a)}{{N-1}}$.

To see this note that by (\ref{line_dynamics}) for  $0<n<N-1$ we have at steady state:
\beq
x_n^{\infty}=\frac{1}{2} \left(x_{n-1}^{\infty}+x_{n+1}^{\infty}\right).
\eeq
 $x_n^{\infty}=a+\frac{n(b-a)}{{N-1}}$ satisfies the asymptotic equation and by (\ref{eq:asymptotic}) there is a unique stable point which is the linear function.

\begin{figure}[htb]
\centering{\includegraphics[width = .4\textwidth]{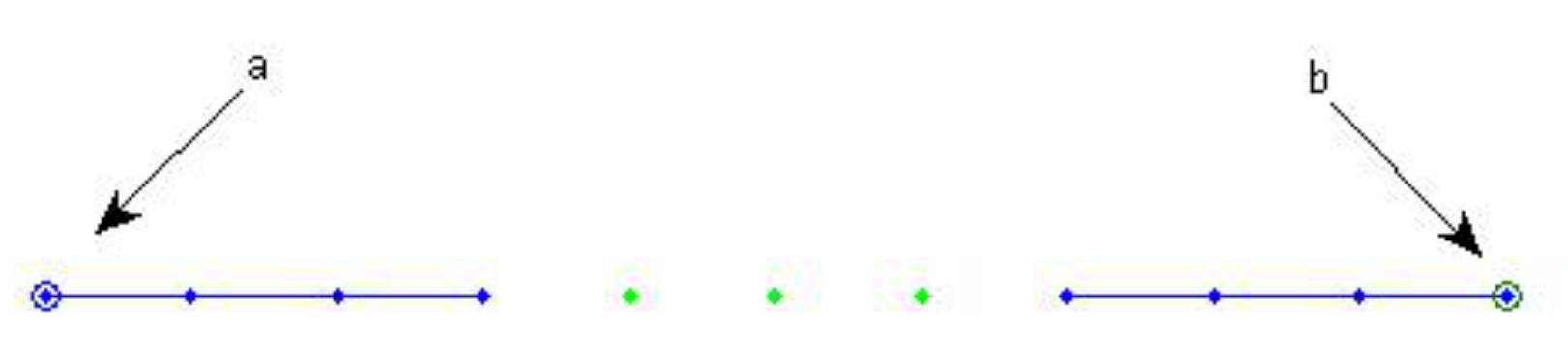}}
\caption{Linear network with boundary conditions.}
\label{line_graph}
\end{figure}

\end{example}

\section{Randomized gossip algorithms}\label{gossip}
This section extends the results of Section~\ref{s:main_theorem} to the case of random gossip algorithms. Here, two types of network operation, sensor polling and pair-wise averaging, are considered.

{\em Sensor polling:} Each sensor decides to poll all its neighbors with probability $p$. Once it decided to poll the neighbours, it computes its new state by weighted averaging with fixed weights given by the matrix $\mL$ (for example based on the degree or any other predetermined set of weights). This is done independently for each sensor.  For simplification purposes, it is assumed that sensors  update their measurement synchronously.
In the sensor polling, each sensor averages its neighbors (applies a row of $\mL$ to the data) with probability $p$ and otherwise, does not change its value (applies a row of the identity matrix) with probability $1-p$. Therefore, we obtain
\beq
\E\left(\mL(k)\right)= p \mL +(1-p) \mI,
\eeq
where $\E(\cdot)$ denotes the expectation. Since sensors act independently over time (the choices are independent between sensors and times), the following expression is obtained for value of each node at time $k$:
\beq
\E\left(\vx(k)\right)=\left(p \mL +(1-p) \mI\right)^k \vx(0)
\eeq
By the convexity of the stochastic matrices,  $\E\left(L(k)\right)$ is stochastic and the sub-block
\beq
\E\left(\mP_i(k)\right)=p \mP_i +(1-p) \mI,
\eeq
is also weakly sub-stochastic.
Following the same argument as in Section~\ref{main_theorem}, we obtain that with probability~1
\beq
\bea{lcl}
\E\left(\vx_\infty(\vx_b(0))\right)&=&\lim_{k \rarrow \infty} \E\left(\vx(k)\right)\\
                                  &=& \left[\vx_b^T, \left( \left(\mI-\mP_i \right)^{-1} \mP_e\vx_b\right)^T \right].
\ena
\label{eq:gossip_const}
\eeq
Similar to Theorems~\ref{main_theorem} and~\ref{thm:constant}, when the boundary conditions are constant, the network converges in the mean to the same constant, and generally with a given boundary conditions, the network converges to the harmonic function defined by (\ref{eq:gossip_const}).

{\em Pair-wise averaging:} The second model is the pairwise interaction model where the matrix $\mL$ is derived from a doubly stochastic matrix. In this model, two internal nodes choose to update their values by averaging, while whenever an interaction occurs between external node and a boundary node, only the internal node updates its value; that is, the value of boundary nodes always remains fixed.
The process can be viewed as follows: We define a probability distribution over pairs $1 \le i,j \le N$  (where $N = M+K$) given by $\pi_{i,j}$ and an averaging constant $0<\ga_{i,j} <1$. For each $i \neq j$, a weighting matrix $\mW_{i,j}$ is defined as
\beq
\left[\mW_{i,j}\right]_{m,n}=\left\{
\bea{lcl}
\gd (m-n)& \hbox{if} & \{m,n\} \not \subseteq \{i,j\} \\
\ga_{i,j}&  \hbox{if} & \{m,n\}=\{i,j\}\\
1-\ga_{i,j}&  \hbox{if} & m=n= (i \hbox{\ or \ }j),
\ena
\right.
\eeq
where $\gd(\cdot)$ is Dirac delta function. At each step a pair is chosen randomly according to $\pi_{i,j}$ and the matrix $\mW_{i,j}$ is applied to the data, i.e. nodes $i,j$ exchange their data and average it with weights given by $\ga_{i,j}, 1-\ga_{i,j}$, with the exception that boundary nodes do not perform the averaging but maintain their original data.
By the convexity of the doubly stochastic matrices
\beq
\E\left(\mW(k)\right)=\sum_{i \neq j} \pi_{i,j} \mW_{i,j},
\eeq
is also doubly stochastic.
Furthermore, the operation of the system is given by a random matrix of the form
\beq\label{eq:Lpairwise}
\mL(k) = \left[
\bea{c|c}
\mI_{K\times K} & {\bf 0}_{K\times M} \\
\hline \\
\mW_e(k) & \mW_i(k)
\ena
\right].
\eeq
 where the lower part of the matrix is a random matrix which consists of the lower $M = N-K$, (recall that $M$ denotes the number of internal nodes) rows of $\mW(k)$.
 Using the same argument as before and assuming that there exists at least one pair $(i,j)$ with $\pi_{i,j}>0$ where $i$ is a boundary node and $j$ is an internal node, we  obtain that
 \beq
 \mP_e = \E\left(\mW_e(k)\right)\neq \vzero
 \eeq
 and
 \beq
\mP_i = \E\left(\mW_i(k)\right)
 \eeq
 is a sub-stochastic matrix. This suffices to prove convergence in the mean:
 \beq\label{eq:pair}
 \lim_{k \rarrow \infty} \E\left(\vx(k)\right) = \left[\vx_b^T, \left( \left(\mI-\mP_i \right)^{-1} \mP_e\vx_b\right)^T \right]^T.
 \eeq
\section{Consequences for networks - Experimental Studies on Humans}
\label{consequences}
In this section, the implications of the main results of this paper (which are presented in Section~\ref{s:main_theorem}) on consensus in social and sensor networks are presented. First, the control of beliefs in social networks, which has significant impact on advertisement technology as well as steering opinions in a Bayesian setup, is discussed. We, further, present  the results of an experimental study which verifies the existence of such boundary nodes in social networks. Then,  an example is provided where a single malicious node can completely destroy the detection capability of a sensor network with distributed averaging.
\subsection{Implications to social networks}\label{sec:SNs}
 Here, an example is provided to discuss the importance of the boundary value problems in the asymptotic agreement of agents  in social networks. The propagation of beliefs and learning in non-Bayesian social networks is investigated in \cite{SL}. One protocol to experiment learning in social networks is consensus formation over graphs \cite{consensus, diffusion}. In this subsection, we investigate the effect of the boundary value problems in such consensus networks. Consider a social network where individuals (agents) interact to update their belief about an underlying state of nature simply by evaluating the weighted average of the beliefs of their neighbors at previous time-instant. Assume that there exists two types of agents in the social network: (i) ordinary agents (internal nodes) whose beliefs change according to the beliefs of their neighbors, and (ii) ``advertiser"s (boundary nodes). The belief of advertisers are not affected by their neighbors, thus, remain fixed over time.  These advertisers can be viewed as boundary nodes that inject  malignant beliefs into the social network. Such a problem, the effect of advertisers (they can also be considered as malicious agents) on the consensus formation in social networks is not investigated in the literature. Here, we can simply show that the belief dynamics in such social networks can be modeled with (\ref{boundary_TI}). Consequently, the asymptotic agreement of agents in social network  can be computed from (\ref{eq:asymptotic}). This means that the asymptotic agreement of the network, no matter of the initial beliefs of the internal nodes, depends only on the malignant beliefs that are being injected into the network by the advertisers. From this, the asymptotic agreement of social network can be controlled and shaped by choosing proper values for the (initial) belief of the advertisers $\vx_b(0)$ in the presence of adverting agents (boundary conditions).

 \subsection{Existence of boundary nodes in human social networks: An experimental study}\label{experimental} Individuals in a society who are not affected by their neighbors  (they stand firm on their own belief) can be considered as boundary nodes in that society.
 In collaboration with the Department of Psychology, University of British Columbia (UBC), we conducted an experimental  study of the existence of such boundary nodes on a group of undergraduate students at UBC during the period October-November 2013. Below we report on these experimental results\footnote{We acknowledge Prof. Alan Kingstone and Dr. Grayden Solman of the Department of Psychology, University of British Columbia, for conducting the psychology experiment.}.

\paragraph*{Experimental Setup}  The experimental study involved 1658 individual experiments. Each individual experiment comprised two participants.
The two participants were asked to perform a perceptual task interactively.  Two arrays of circles were given to each pair of participants, then, they were asked to judge which array had the larger average diameter. One member was chosen randomly and started the experiment and chose either left side or right side as his judgment. Thereafter, each member saw their partner's previous response and his own previous judgment prior to making their own judgment; thereby providing a means for measuring social influence. The participants continued choosing actions according to this procedure until the  experiment terminated. The experiment terminated
when the response of each of the two participants did not change for three successive iterations (the two participants did not necessarily have to agree
for the experiment to terminate). In this experimental study, each participant chose an action  $a \in \mathbf{A} = \{0,1\}$;  $a = 0$  when he judged that the left array of circles had the larger diameter and $a = 1$ when his judgments  was that the right array of circles had the larger diameter. In each experiment, judgments (actions) of participants are recorded along with the amount of time taken by each participant to make that judgment.

\begin{figure*}[!t]
\centerline{
\includegraphics[width=.7\textwidth]{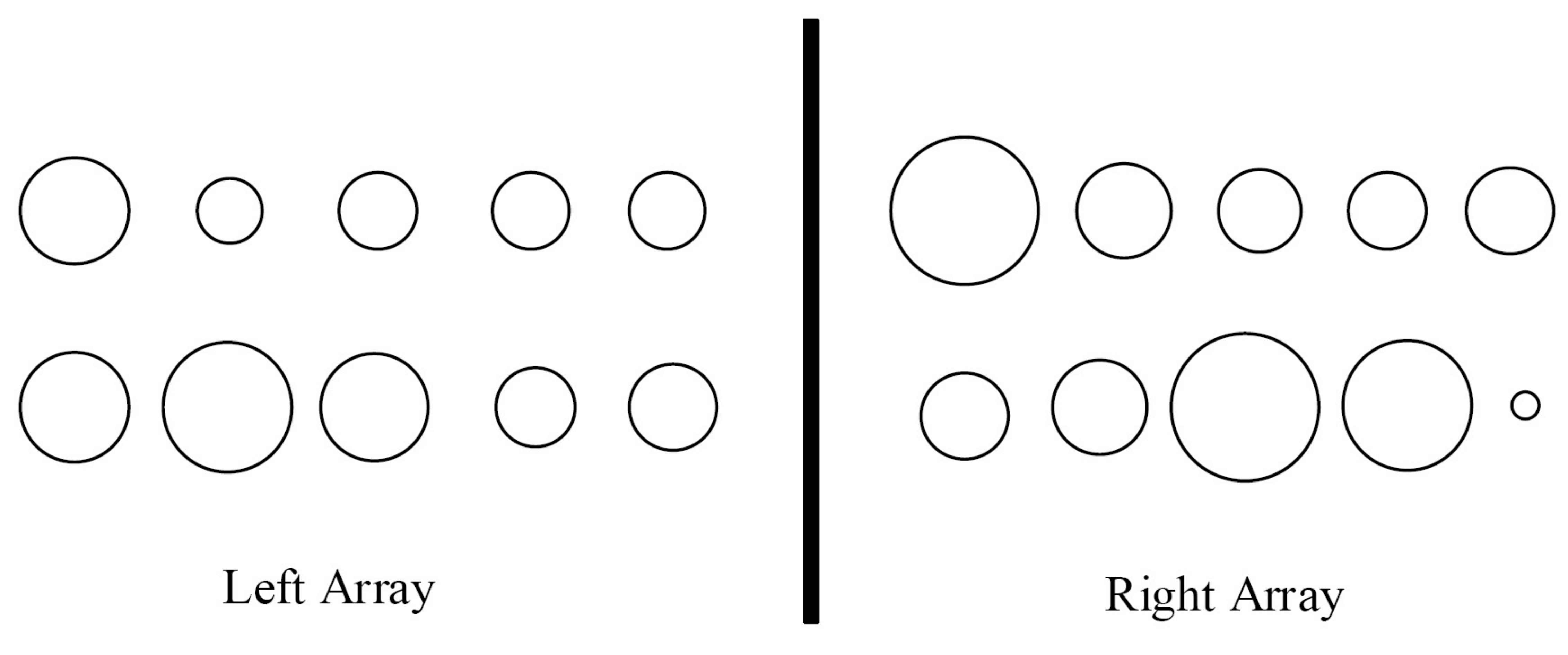}}
\caption{Two arrays of circles were given to each pair of participants. Their task is to interactively determine which side (either left or right) had the larger average diameter. In this example, the average diameter of the left array of circles is 8.4 mm and the right array is 9.1 mm.}
\label{Fig:SocialSensor}
\end{figure*}

\begin{figure}[htb]
\centerline{
\includegraphics[width=.5\textwidth]{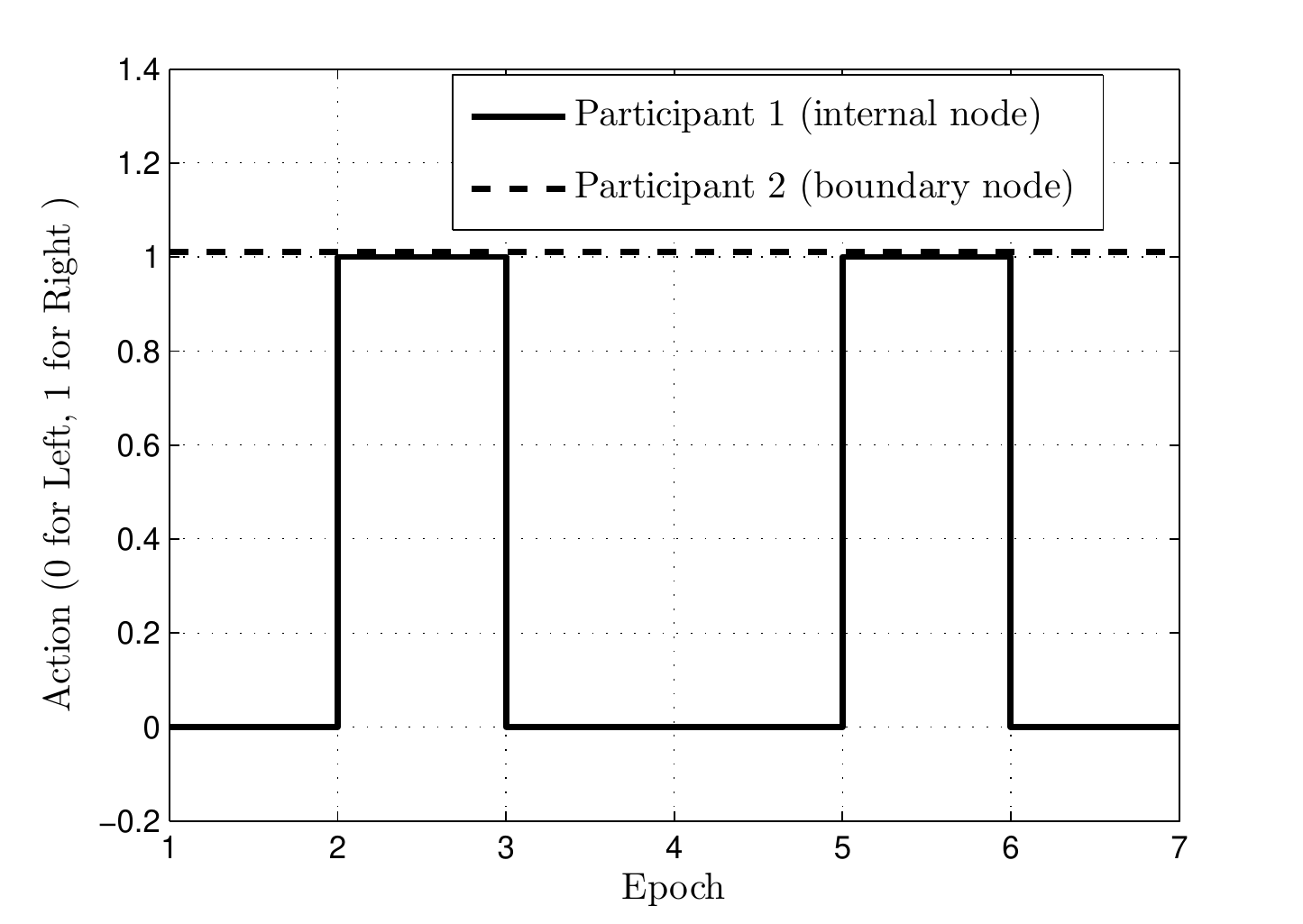}}
\caption{Actions of two participants in a group at different epochs. Participant~1 can be considered as an internal node and Participant~2 can be viewed as a boundary node.}
\label{Fig:Samplepath}
\end{figure}

 \paragraph*{Experiment Results} Surprisingly, among 3316 individuals who participated in this experiment, 1336 participants (around 40\%) did not change their judgements after observing the action of their partners (these participants can be viewed as boundary nodes), while the other 60\% changed their initial judgment and got influenced by the action of their partners (internal nodes). Fig.~\ref{Fig:Samplepath} shows the sample path of two participant in a same group, Participant~1  is an internal node while Participant~2 is a boundary node.

 We, further, investigate the time taken by each participant to make his judgment. Let $\mu_{\rm judg.}$ and $\sigma_{\rm judg.}$ denote the mean and the standard deviation of the time taken by participants to make their judgments in milliseconds. The results of our experimental study, which are presented in Table~I, show that the internal nodes,  in average, require more time to make their judgments compared to the boundary nodes; this is quite intuitive from the fact that the boundary nodes stand firm on their decisions and ignore the judgment of their partners and thus require less time to make their judgments.

\begin{table}[h]
\begin{center}
 \begin{tabular}{ | c || c | l | l |  }
 \hline
  Type of nodes & relative frequency & $\mu_{\rm judg.}$ & $\sigma_{\rm judg.}$ \\ \hline
  Internal & 40 \% & 1058 ms & 315 ms \\ \hline
  Boundary & 60 \% & 861 ms & 403 ms\\ \hline
\end{tabular}
 \end{center}
 \label{table1}
 \caption{The frequency of the internal and the boundary nodes in a community of 3316 undergraduate students of the University of British Columbia along with the statistics of the time required by participants (of both types) to make their judgments in milliseconds.}
 \end{table}
 This psychology experiment illustrates the importance of investigating the consensus of networks with boundary nodes in social networks.

\subsection{Implications to sensor networks}
 In this subsection, an example is provided which shows the sensitivity of consensus networks to malicious attacks, and even for an unintended bias in the computation of the mean in one of the elements of the network. To observe the effect, consider a binary distributed detection problem. Two hypotheses $H_0,H_1$ are given and each sensor $0 \le n \le N-1$ measures a realization $x(n)$
\beq
x(n)\sim\left\{
\bea{ll}
P\left(x(n)|H_0\right) & H_0 \\
P\left(x(n)|H_1\right) & H_1
\ena \right.
\eeq
Here, $P(\cdot)$ denote the probability of an event. Assume that $x(n)$ are conditionally independent given $H_0$ and $ H_1$. To obtain an optimal decision, each sensor needs to compute the average log-likelihood over the network.
The log likelihood ratio is given by:
\beq
L(\vx)=\log \frac{P\left(x(0),...,x(N-1)|H_1\right)}{P\left(x(0),...,x(N-1)|H_0\right)}
\eeq
which can be computed by conditional independence as
\beq
L(\vx)=\sum_{n=0}^{N-1} \log \frac{P\left(x(n)|H_1\right)}{P\left(x(n)|H_0\right)}
\eeq
This detector can be easily computed using a gossip algorithm. However given any threshold designed for a fixed probability of false alarm $\gt$,
a malicious node, that is aware of the desired detection threshold $\gt$, can choose a value $\gm<\gt$ and decrease the probability of detection to $0$ by steering the network to $\gm$.
 The system matrix is now given by
\beq\label{eq:A2}
\mL = \left[
\bea{c|c}
1 & {0} \\
\hline \\
\gm\ve_1 & \mP_i
\ena
\right],
\eeq
where  index $1$ is used to denote the faulty node.
By Theorem~\ref{thm:constant}, the limiting value at the network nodes, given by
\beq
\vx_{\infty}=\gm \vone_N<\gt,
\eeq
is constant. Once the network converged to consensus, all nodes will agree that there is no target.
The level of confidence can be made arbitrarily high by choosing smaller $\gm$.

Note that the existence of a consistent bias in a single node can result in similar erroneous conclusion, or if the bias is larger than the detection threshold, it will generate false alarms.

\section{Networks with time periodic boundary conditions}
\label{periodic}
In this section, the analysis of Section~\ref{s:main_theorem} is extended  to the case where the boundary conditions are periodic with period $\gt$.
Similar to Theorem~\ref{main_theorem},  consider the following system:
Let $\gt>0$ be fixed.
\beq
\bea{ll}
\hbox{Initial conditions: } & \\
&\hspace{-1cm} \vx_i(0)=[x_{K+1}(0),...,x_{M+K}(0)]^T \\
\hbox{Boundary conditions: } &  \\
\forall t \le \gt-1,n: &\hspace{-1cm} \vx_b(t+n\gt)=[x_1(t),...,x_K(t)]^T, \\
\hbox{System dynamics: }\\
\forall t > 0: & \hspace{-1cm} \vx_i(t+1)=\mP_e \vx_b(t) + \mP_i\vx_i(t).
\ena
\label{boundary_periodic}
\eeq
The matrix $\left[\mP_e, \mP_i \right]$ is stochastic, which implies that $\mP_i$  is weakly sub-stochastic. As a consequence, $\gr(\mP_i)<1$.
Similar to the time invariant case, we can show inductively the following theorem:
\bthm
The state of the internal system $\vx_i(k)$ at time $k$ is given by
\beq
\label{finite_periodic}
\vx_i(k)=\sum_{m=0}^{k-1} \mP_i^m \mP_e \vx_b(k-m \mod \gt)+\mP_i^k \vx_i(0)
\eeq
Furthermore,
\beq
\label{lim_periodic}
\lim_{k\rarrow \infty} \vx_i(k) = \sum_{m=0}^{\gt-1} \left(\mI-\mP_i^\gt\right)^{-1} \mP_i^m \mP_e \vx_b(m).
\eeq
\ethm
 \begin{proof} The proof of (\ref{finite_periodic}) is an easy induction. To show the second part, one splits the sequence in (\ref{finite_periodic}) according to the value of $m \mod \gt$, and noting that we have a geometric series with factor $\mP_i^\gt$ for each subsequence, beginning with $\mP_i^m \mP_e \vx_b(m)$. Since $\gr(\mP_i)<1$ the part depending on initial conditions in the interior part of the network tends to $\bf 0$ exponentially fast. \end{proof}

\section{Numerical Examples}
\label{simulations}
\begin{figure}[htb]
\centering{\includegraphics[width = .5\textwidth]{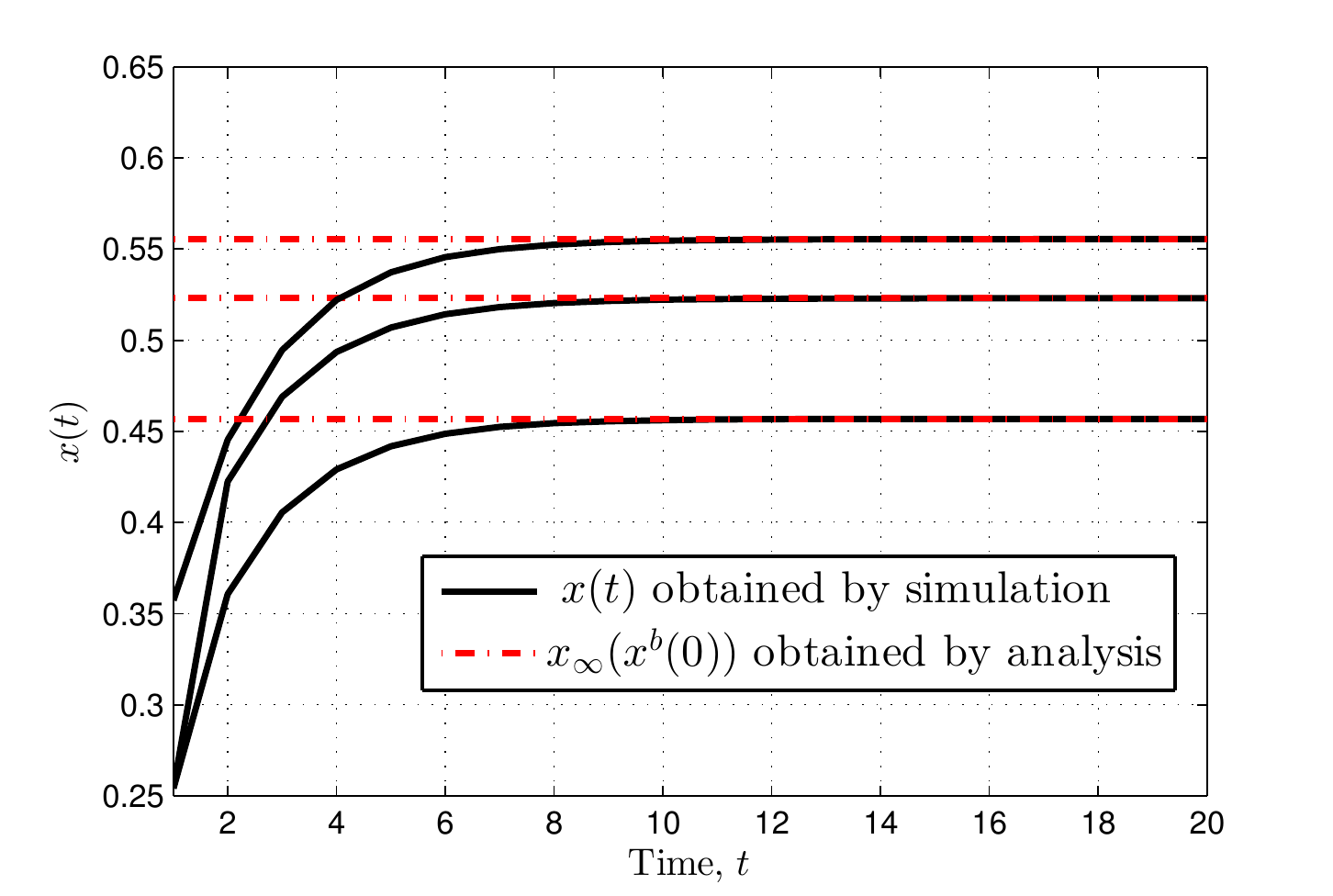}}
\caption{Dynamics of beliefs in a network with boundary constraints.}
\label{fig1}
\end{figure}

In this section, numerical examples are provided to verify the results of Sec.\ref{gossip} and Sec.\ref{consequences}. In the first numerical study, we investigate the dynamics of consensus averaging in networks with boundary constraints. A network comprising of $N = 100$ nodes ($K = 50$ boundary nodes and $M = 50$ internal nodes) is considered. In this network, internal nodes (in contrast to boundary nodes whose values remain fixed) average over their neighbors to update their values.   Neighbors of each node and  weights used in the averaging (that is matrix $\mL$) are chosen randomly. The initial values of nodes $\vx(0)$ are simulated from normal distribution; that is $\vx_i(0) \sim \mathbf{N}(0,5)$ and $\vx_b(0) \sim \mathbf{N}(0,1)$.  Fig.\ref{fig1} compares $\vx_\infty(\vx_b(0))$ obtained via (\ref{main_eq}) in Theorem~\ref{main_theorem} with $\vx(t)$ from simulation. Three different entries of $\vx_\infty(\vx_b(0))$ and $\vx(t)$ are shown in Fig.\ref{fig1}.

\begin{figure}[htb]
\centering{\includegraphics[width=.5\textwidth]{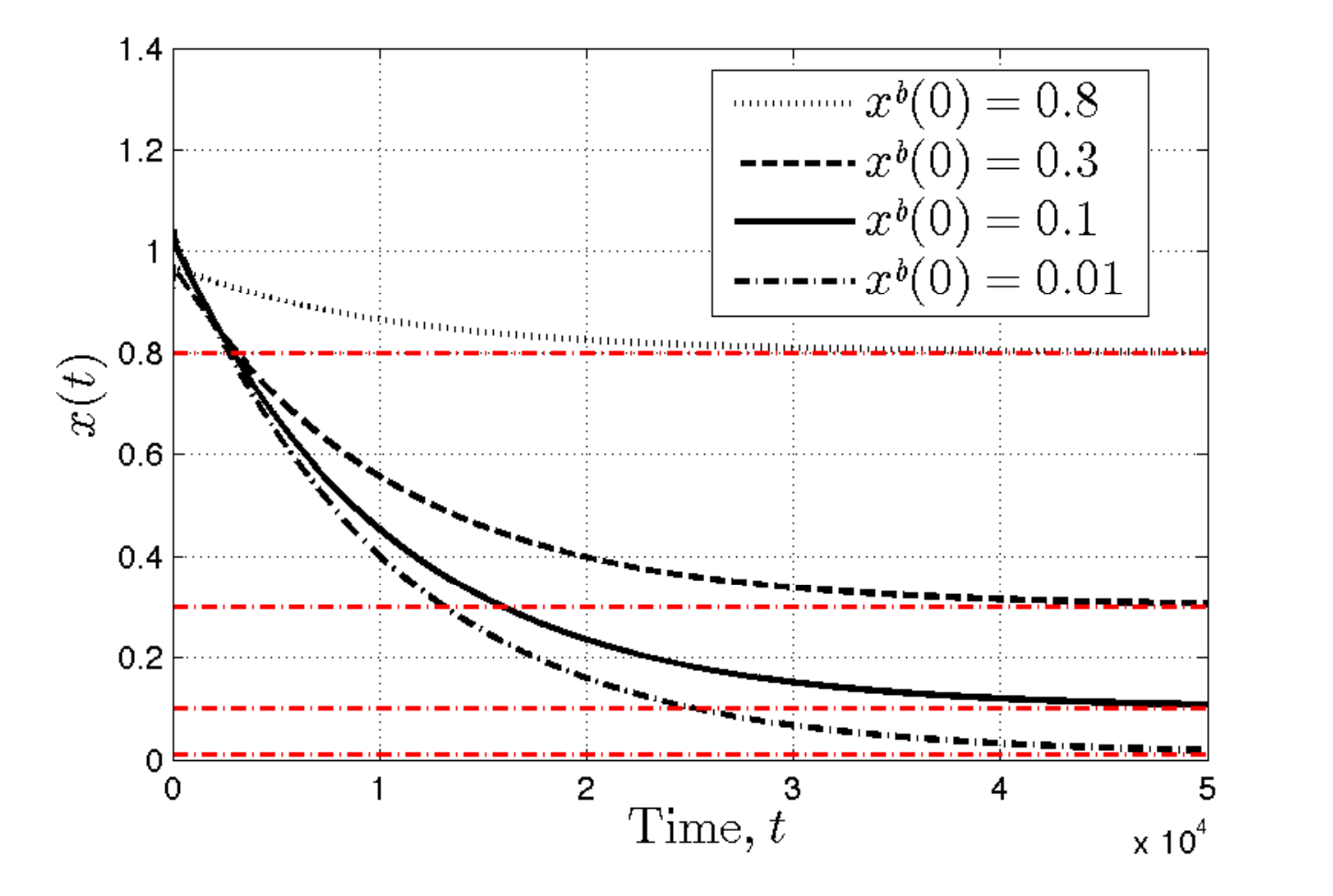}}
\caption{ Effect of an advertising node (boundary node) in the asymptotic belief of internal nodes.}
\label{fig2}
\end{figure}

In the next example, we study the effect of advertisers (boundary nodes) in formation of consensus in a network. A network with $N = 50$ nodes ($K = 1$ boundary node) is considered. In this scenario, the advertising node (boundary node) injects a fixed belief into the network (that is, it is not affected by belief of its neighbors) and the internal nodes average over their neighbors to update their beliefs. Similar to the previous example, the neighbors and the averaging weights are chosen randomly. Fig.~\ref{fig2} depicts $\vx(t)$ at each iteration for one of the non-advertiser (internal) agents for different values of initial belief $\vx_b(0)$. As can be seen in Fig.~\ref{fig2}, the beliefs of internal nodes converge to the initial belief of the advertising node (boundary node). This is quite interesting result in social networks, it shows that by the means of advertising node in a social network, we can control the consensus of the whole network.

\begin{figure}[htb]
\centering{\includegraphics[width=.5\textwidth]{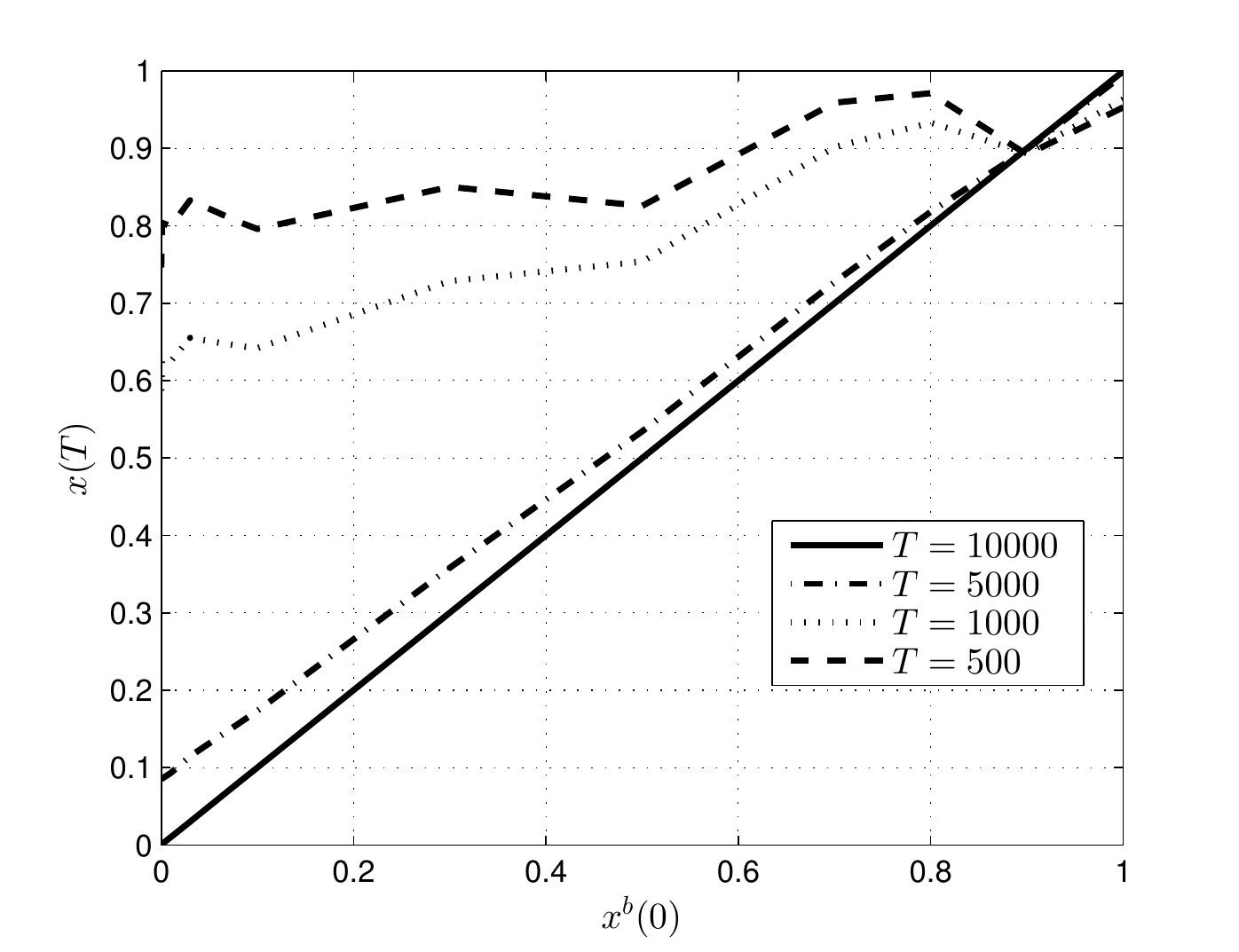}}
\caption{Belief of an internal node at time $T$ versus the initial belief of the advertising node $\vx_b(0)$.}
\label{fig4}
\end{figure}

To study the rate of convergence of beliefs to the asymptotic value, a network similar to the previous example with $N = 50$ nodes and $K = 1$ boundary node is considered. Fig.\ref{fig4} shows the value of $\vx(T)$ for one of the internal nodes versus the initial belief of the advertising node $\vx_b(0)$ for different values of $T$. For sufficiently large values of $T$ (for example $T = 10000$), belief of the internal node reaches its asymptotic value (which is equal to $\vx_b(0)$), therefore, $\vx(T) \approx \vx_b(0)$; the slope of the solid line which corresponds to $T= 10000$ is very close to one. Fig.~\ref{fig4} reveals that for smaller values of $T$, beliefs of the internal nodes converge faster to their asymptotic values when the initial belief of the advertising node $\vx_b(0)$ is larger.

\begin{figure}[htb]
\centering{\includegraphics[width=.5\textwidth]{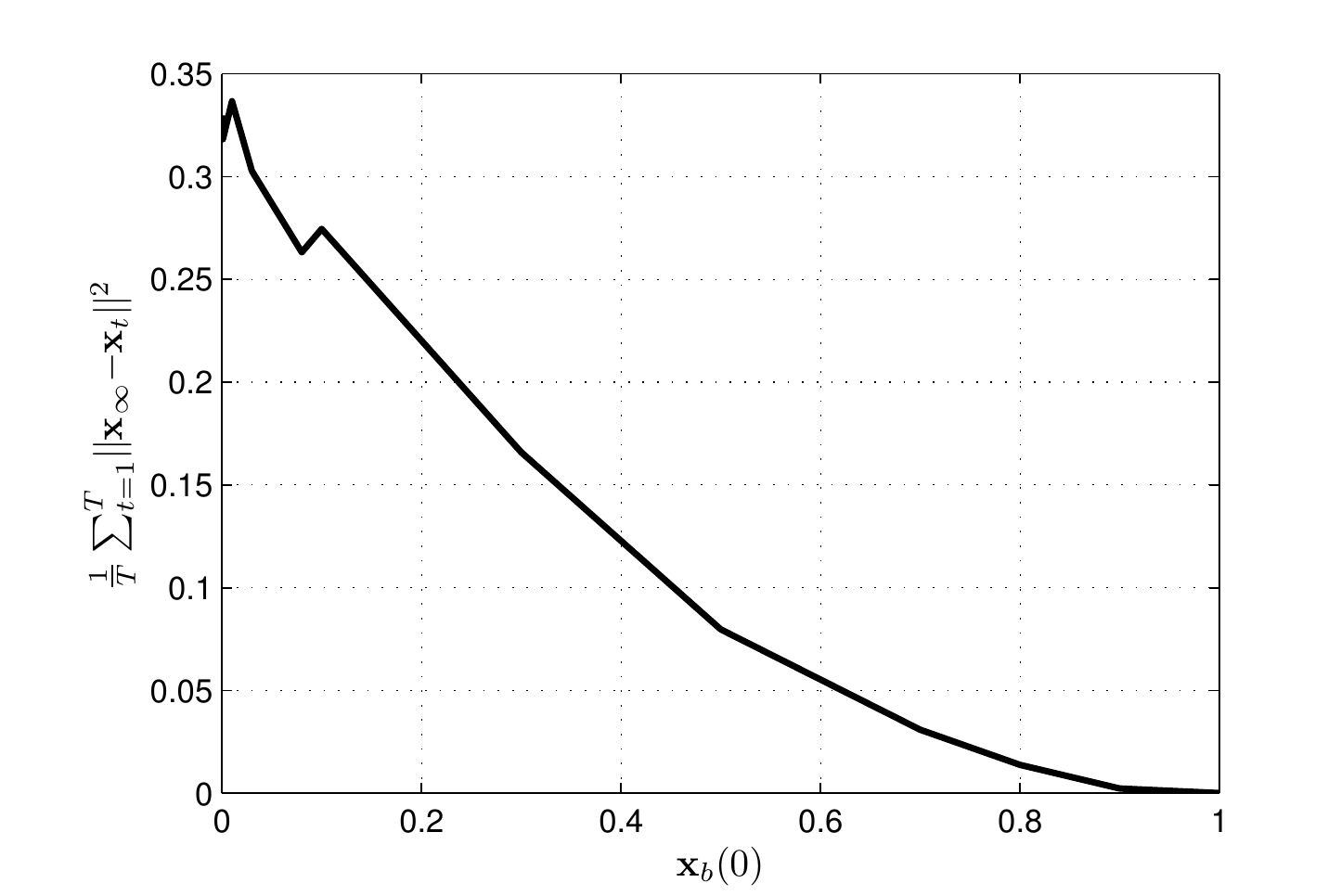}}
\caption{The average distance between beliefs of an internal node at different times and the consensus of the network. }
\label{fig3}
\end{figure}

Fig.\ref{fig3} illustrates the average of the distance between $\vx_\infty(\vx_b(0))$ and $\vx(t)$ for $t = 1,2,\ldots, 2000$   versus the initial belief of the advertising nodes $\vx_b(0)$. Similar to the previous example, it can be inferred from Fig.\ref{fig3} that the distance between beliefs of internal nodes and the initial belief of the advertising node is lower when $\vx_b(0)$ is larger. In other words, the belief of internal nodes converge faster to the asymptotic value when the bias (the belief of the advertising node $\vx_b(0)$) is large enough.

\begin{figure}[h]
\centerline{
\includegraphics[width=.5\textwidth]{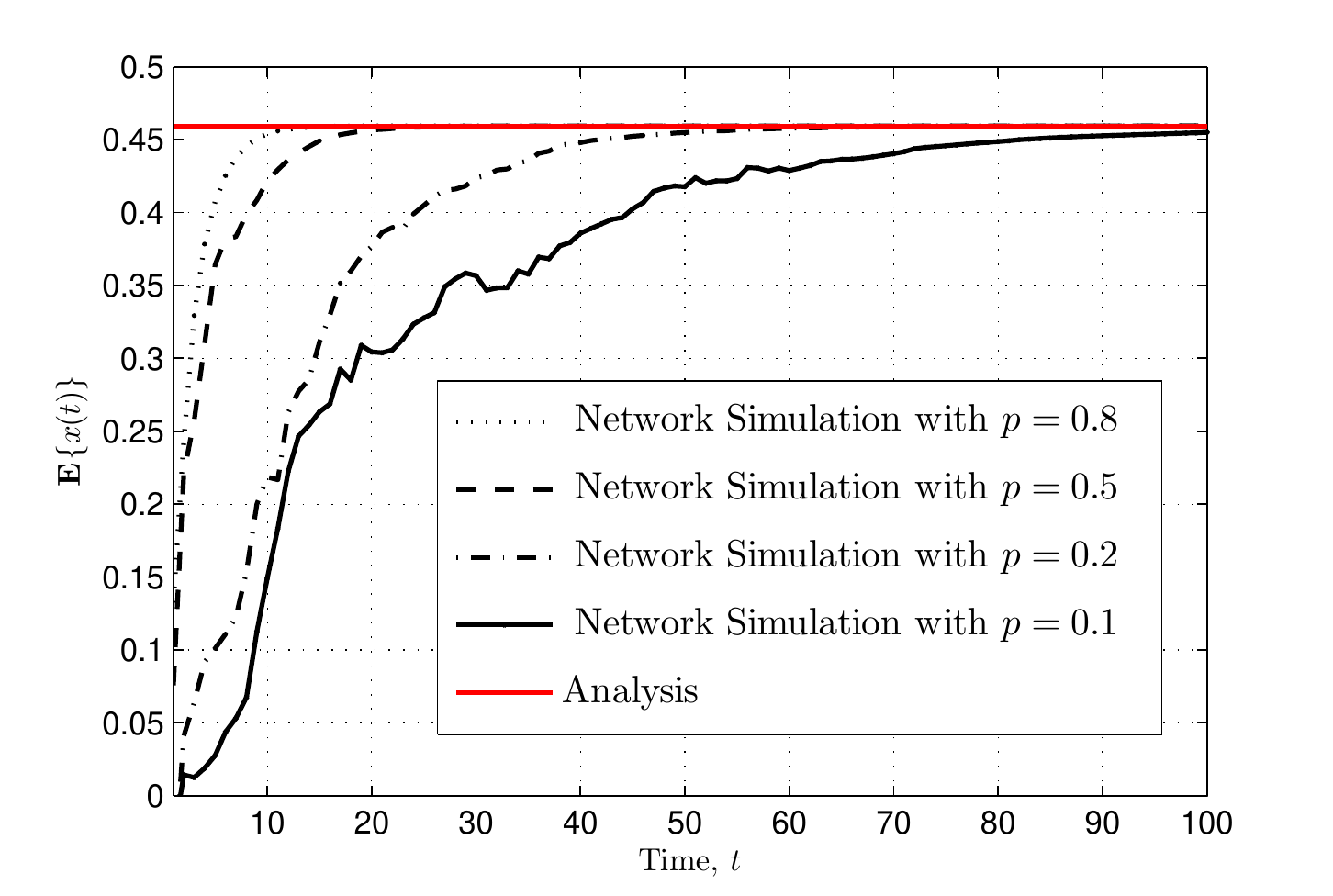}}
\caption{Beliefs of an internal node in a network with sensor polling in the existence of boundary constraints for different probabilities of polling.}
\label{Fig:gossip}
\end{figure}

Here, we study the effect of the boundary value constraints in random gossip algorithms.  A network comprising of $N = 100$ sensors ($K = 50$ boundary nodes and $M = 50$ internal node) is considered. In this scenario, each sensor independently decides to poll its neighbors with probability $p$; it averages over the neighbors to update its belief. In other words, with the probability $1 - p$ its belief remains unchanged. Neighbors of each sensor and the weights used in averaging; that is matrix $\mL$ ($\mP_e$ and $\mP_i$) is chosen randomly. The initial values for beliefs $\vx$ are simulated from normal distribution; that is $\vx_i(0) \sim \mathbf{N}(0,5)$ and $\vx_b(0) \sim \mathbf{N}(0,1)$. This scenario is simulated $1000$ times and the average of an internal sensor's beliefs $E(\vx_\infty(\vx_b(0)))$ for four different values of $p$, ($p \in \{0.1, 0.2, 0.5, 0.8\}$) are depicted in Fig.~\ref{Fig:gossip} along with the expected consensus of the network computed via (\ref{eq:gossip_const}). As can be seen in Fig.~\ref{Fig:gossip}, the network converges to the harmonic function defined by (\ref{eq:gossip_const}) in the existence of boundary constraints.

\begin{figure}[h]
\centerline{
\includegraphics[width=.5\textwidth]{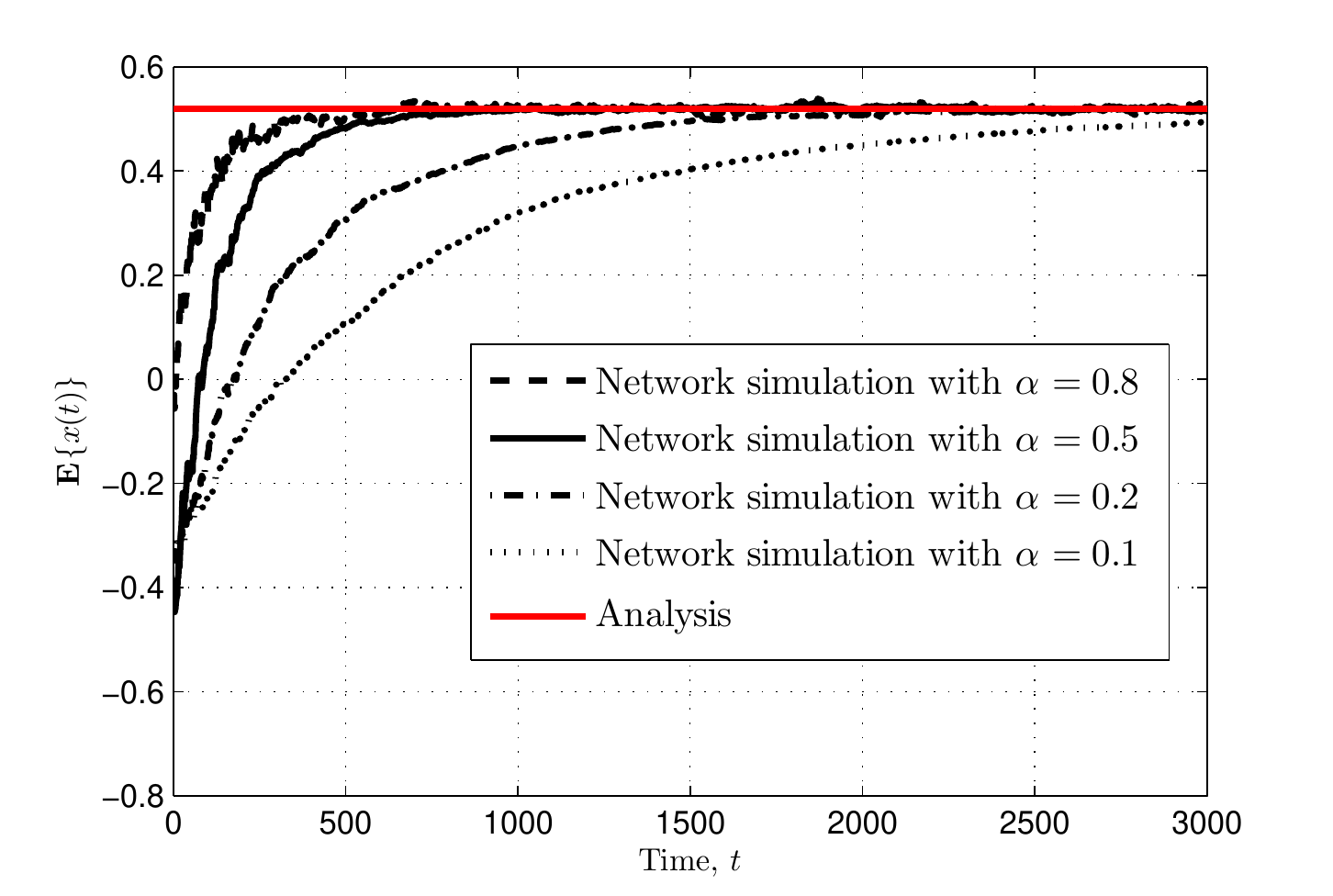}}
\caption{Beliefs of an internal node in a network with pairwise interaction model between nodes in the existence of boundary constraints for different averaging weights.}
\label{Fig:pair}
\end{figure}

In the next example, a pairwise interaction model for the network is considered.  Similar to the previous example, a network comprising of $N = 100$ nodes ($K = 50$ boundary nodes and $M = 50$ internal node) is considered. At each time, two nodes are randomly selected according to a uniform probability distribution over pairs. Then, these nodes (given that none of them are boundary nodes) update their beliefs by averaging. In the interaction between an internal node and a boundary node, only the internal node updates its belief, and in the interaction between two boundary nodes, none of them update their beliefs. We assume that for all pairs of nodes, the averaging weight is fixed, i.e., $\alpha_{i,j} = \alpha$ for $ 1 \leq i,j \leq N$ and $0< \alpha < 1$. The initial values for beliefs $\vx$ are simulated from normal distribution; that is $\vx_i(0) \sim \mathbf{N}(0,5)$ and $\vx_b(0) \sim \mathbf{N}(0,1)$. This scenario is simulated $1000$ times and we average over beliefs of one the internal sensors. $E(\vx_\infty(\vx_b(0)))$ for four different values of $\alpha$, ($ \alpha \in \{0.1, 0.2, 0.5, 0.8\}$) along with the expected consensus of the network computed via (\ref{eq:pair}) are depicted in Fig.\ref{Fig:pair}. As can be seen in Fig.~\ref{Fig:pair}, the network converges in the mean to the asymptotic value defined by (\ref{eq:pair}) in the existence of boundary constraints.

\section{Conclusions}
This paper has examined the effect of boundary valued conditions on consensus. The main result
was Theorem~\ref{main_theorem} which gave sufficient conditions for the
consensus only to depend on the boundary nodes and not the internal nodes. Applications in sensor polling and human interactions (with experimental data) were described.
Numerical examples were provided to illustrate the main result.

\section*{Acknowledgment}
We would like to acknowledge our colleagues in the Department of Psychology, University of British Columbia, Prof. Alan Kingstone and Dr. Grayden Solman, for conducting the psychology experiment discussed in Sec.\ref{sec:SNs}.
\bibliographystyle{IEEEtran}

\end{document}